\documentclass[12pt]{article}

\usepackage{graphicx}

\title{The two-channel exotic charmonium-like resonances in the mass region $(3900-4700)$ MeV}
\author{A.M.Badalian and  Yu.A.Simonov \\
NRC ``Kurchatov Institute'' \\
Moscow, Russia}

\newcommand{\beq}{\begin{eqnarray}}
 \newcommand{\eeq}{\end{eqnarray}}
\newcommand{\be}{\begin{equation}}
 \newcommand{\ee}{\end{equation}}

 \def\la{\mathrel{\mathpalette\fun <}}

\def\fun#1#2{\lower3.6pt\vbox{\baselineskip0pt\lineskip.9pt
\ialign{$\mathsurround=0pt#1\hfil ##\hfil$\crcr#2\crcr\sim\crcr}}}

\newcommand{{\SD}}{\rm SD}

\newcommand{{\Mc}}{\mathcal{M}}

\newcommand{\vep}{\mbox{\boldmath${\rm p}$}}

\begin{document}
\maketitle
\begin{abstract}
Numerous resonances in the $c\bar c$ and $4q$ systems, containing $c\bar c$  plus $s\bar s$ quarks (or light $q\bar q$) were observed  during last decades
and recently the LHCb has found a remarkable narrow peak $T_{cc}(3875)$ in the $D D^*$ system. Besides there are several highly excited  charmonium-like resonances, which  can be treated as the shifted standard charmonium states. We analyze all these groups  of the resonances in the mass region $(3900-4700)$ MeV, using relativistic strong coupling theory
with possible  channel coupling phenomena. For the shifted charmonium states conventional charmonium spectrum is presented, being calculated with the relativistic string Hamiltonian, which does not contain fitting parameters, while for high excitations the universal flattened confining potential is used. It is  shown that $X(4274),X(4500),X(4700)$ can be
identified as $3\,^3P_1, 4\,^3P_0, 5\,^3P_0$ states. The group of exotic states are considered using the Extended Recoupling Model, where two mesons $m_1,m_2$ transfer into another pair of mesons $m_3, m_4$ and back (infinite number of times), creating the four-quark resonance. Within this approach the resonances  -- $T_{cc}(3875), Z_c(3900), X(3915), Z_{cs}(3985), X(4014)$, and $X(4140)$ -- can be explained as the exotic four-quark states in the $S$-wave decay channels.
\end{abstract}

\section{Introduction}

Recently new charmonium-like resonances have been found by several groups \cite{1,2,2*,3,4,5,5*}, e.g.  $X(4685), X(4740)$  was observed by the LHCb in
the $B^+\rightarrow J/\psi \phi$ and $B^+\rightarrow J/\psi K^+$ decay channels \cite{2,2*}, while  narrow $Z_{cs}(3985)$ structure was discovered
by the BESIII in the $e^+e^-\rightarrow K^+(D_s^-D^{*0} + D_s^{*-}D^0)$ reaction \cite{3}. Together with previously reported resonances, $X(4140),$ $X(4500),$ $X(4700)$ \cite{4} and $X(3915)$, observed by several groups \cite{5}, now one has numerous examples of the $D\bar D$ type of resonances near corresponding thresholds with the width in the range from $10$ to $100$ MeV.
At the same time recently the LHCb group was discovered the $T_{cc}(3875)$ resonance in the $DD^*$ system with extremely small
width of 410 keV  \cite{6}, while the search for the resonances in the systems $D_sD_s, D_s^*D_S^*$ has no given  results \cite{6*}. One may wonder about the dynamics of the four quark (two hadron) systems which can give such variety  of resonances. The main purpose of our paper is to describe the physics of exotic states in the $c\bar c$, $cc\bar q\bar q$, $c\bar c q\bar q$
systems, using the standard and new nonperturbative methods in QCD. One can distinguish two types of  resonances in such systems:
first, new type of the $qq\bar q\bar q$ systems with nontrivial dynamics and the standard $q\bar q$ systems, strongly shifted  due to channel coupling with other hadron states. Here these states
will be considered in the region $(3850-4700)$ MeV.

The case of the $T_{cc}(3875)$ resonance, recently found by the LHCb in the $D^0D^0\pi^+$ mass spectrum \cite{6} near the $D^*D$ threshold ($M=3875$ MeV), looks to be very surprising, since it has very small width of 410 KeV. Such resonance in the $D^*D$ system with very small width has dynamics which is seemingly very different from that in the $D^*\bar D$ resonances with larger width \cite{1,2,3,4,5,5*}. Then the question arises what is the internal dynamics of these resonances and at this point one must find the mechanisms ensuring the resonance characteristics  both in the $D^*D$ and $D^*\bar D$ systems. At the same time no resonances were found in the double-heavy-light systems: $D_s D_s$ and $D_s*D_s^*$ \cite{6*}. It is clear that new
theory must explain all these three facts. This kind of theory was recently suggested in \cite{7} and called  the Extended Recoupling Model (ERM), which develops the previous Recouplimg Model (RM),  suggested in \cite{7*}. As was shown in \cite{7}, the ERM allows to explain the resonances in the systems $DD^*$ and $D\bar D^*$ and the absence of the resonances in $D_sD_s,D_s^*,D_s^*$.
Here to describe four-quark states we use the mechanism of the extended recoupling model (ERM) \cite{7}, where it is assumed that the system of two mesons, e.g. ($J/\psi+\phi$), can transfer into another pair of the mesons by rearranging confining strings and back in the infinite chain, like
$J/\psi \phi \rightarrow (D_s^+\bar D_s^-) \rightarrow J/\psi \phi \to ...$, or there is another chain: $J/\psi K^{*} \rightarrow D_s \bar D^{*} \rightarrow  (D_s^{*} \bar D) \to ...$ etc. Note that
these sequences could be treated, for example,  in the standard OBE approximation with the meson exchanges, which, however, does not contribute singularities nearby the thresholds.
On the contrary, as shown in \cite{7}, the ERM  accounts for the string rearrangement of the four quark system and, as we shall show below, can produce the singularities just near the thresholds. Of special importance is the ability of the ERM model to predict in this way very narrow resonances with the width $\la 1$ MeV. Here this effect will be obtained for the resonance $T_{cc}(3875)$, being in agreement with experimental data \cite{6}. In the ERM the strong coupling occurs in the $S$-wave hadron-hadron systems and therefore  we apply the RM formalism only to  the $S$-wave decay channels, which produce the resonances with $J^P=0^+,1^+,2^+$, and compare our results with existing experimental data.

In the literature  exotic four-quark resonances have been considered using different mechanisms, including standard coupled channel method \cite{8}, the tetraquark model within the Born--Oppenheimer approach \cite{9,10,11,12}, the cusp phenomenon and the triangle singularity \cite{13,14,15}, the few-body theory of hadron-hadron interaction \cite{16}, the analytic approach including the boson exchange mechanism \cite{17,18,19,19*}, and the QCD sum rule method \cite{19**}.

Here besides exotic near-the-threshold states we consider high resonances in the region $4300-4700$ MeV, which are  shifted with respect to the standard $c\bar c$ positions, using the theory
developed before for different meson states in \cite{20,21,22,23}. For that it is important to identify high excitations of conventional $c\bar c$ states (with $n\geq  3$),
which can exist in the same mass interval. Here we use the analysis of the charmonium spectrum from the Badalian, Bakker paper \cite{20,21}, where the relativistic string Hamiltonian is used \cite{22}. This Hamiltonian was derived in the relativistic Fock--Feynman--Schwinger formalism  \cite{23} and does not contain fitting
parameters, providing  a good agreement with experiment for different meson spectra.

The structure of the paper is as follows. In next section we shortly give the basics of the ERM and define the necessary parameters. In section 3 these parameters are used to predict the
masses of the recoupled four-quark resonances and compare those with experimental data. In section 4 the masses of the $c\bar c$ mesons \cite{20,21} are given and compared with the experiment.   In concluding section the summary of our results are given and the nature of the  $T_{cc}(3875),X(3915),X(3960),Z_{cs}(3985),X(4014)$ resonances in the ERM mechanism is discussed.

\section{The Extended Recoupling Model}

One can study the experimental process where, among other products, specifically two hadrons are produced (the pair 1), which can transfer into another pair of hadrons (the pair 2) with the probability amplitude $V_{12}(\vep_1,\vep_2)$, where $\vep_1,\vep_2$ are relative momenta in two pairs of hadrons.
One can consider an infinite set of transformations, where the hadrons between acts of transformation propagate freely with the
Green's functions $G_i(E,\vep_i),~i=1,2.$ Then the infinite set of transformations from the pair 1 to the pair 2 can be written
as,
\be
G_1 V_{12} G_2 + G_1 V_{12} G_2 V_{21} G_1 V_{12}G_2 + G_1 V_{12} G_2 V{21} G_1 V_{12} G_2 V_{21} G_1 V_{12} G_2 + ...=
= G_1 V_{12} G_2 (1 + N + N^2 + ...)= G_1 V_{12} G_2 \frac{1}{1- N},
\label{1}
\ee,
where $N=G_2 V_{21} G_1 V_{12}$. As a result the total production amplitude $A_2$ of the pair 2 can be written as a product of the slowly varying function $F(E)$ and the possibly singular factor $f_{12}(E)= \frac{1}{1-N}$, so that $A_2= F(E) f_{12}(E)$.

The main point of the whole problem is the transition amplitude $V_{12}= V_{21}$. In  usual approaches one can take it
as one or more OBE diagrams with meson exchanges. In the RM a new mechanism was suggested, where the meson transformation occurs instantaneously due to the recoupling of confining strings between pairs of the mesons. Due to that one obtains a nonfactorizable
specific function of initial and final hadron momenta - $V_{12}(p_1,p_2)$, which may be not suitable for  description of the
general transformation process. In the ERM \cite{7} a more general scheme was suggested through the intermediate stage of the Quark Compound Bag (QCB) type (see \cite{24} for earlier applications),
when all quarks and antiquarks  of two hadrons are participating in the string recoupling (and possibly the spin recoupling) and other energy independent processes. Denoting the QCB wave functions as $\Phi(q_k)$, k=1,2,3,4 and the two-hadron wave functions as $\Psi_i(h_1,h_2)$ , one can write for $V_{12}$:
\be
V_{12}=(\Psi_1(h_{a1}h_{b1})\Phi(q_i))|(\Phi(q_i)\Psi_2(h_{a2}h_{b2})= V_1(\vep_1)V_2(\vep_2).
\label{2}
\ee
For the factor $V_{12}$ it gives the factorized form:  $V_{12}(\vep1,\vep_2)= v_1(\vep_1) v_2(\vep_2)$ and then the amplitude $N$ can be written as $N= z I_1(E) I_2(E)$, where $z=z(E)$ can be called the transition probability and $I_1(E), I_2(E)$ are the following integrals (see \cite{7}):
\be
I_i(E)= v_i G_i v_i= \int{\frac{d^3 p_i}{(2\pi)^3}\frac{v_i^2(p_i)}{E'(p_i) + E^{''}(p_i) - E}},
\label{3}
\ee
where the hadron energies $E'(p_i),E^{''}(p_i)$ in the i-th pair, $E'(p)=\frac{p^2}{2m'}+m'$, define corresponding thresholds $E^{th}_i$ and the reduced masses $\mu_i$, namely,
\be
E^{th}_i=m'(i)+ m^{''}(i), \mu_i= \frac{m'(i)m^{''}(i)}{m'(i)+ m^{''}(i)}.
\label{4}
\ee
The result of integration in $I_i(E)$ can be approximated in the form:
\be
I_i= {\rm const}_i \frac{1}{\nu_i-i\sqrt{2\mu_i(E-E^{th}_i)}},
\label{5}
\ee
where the reduced mass $\mu_i$ was defined above and  $\nu_i $ is expressed via the parameters of the hadron wave functions, participating in the given stage of the recoupling transformation -- see explicit equations in \cite{7} and Appendix of the present paper.
One can see that $I_i(E)$ looks like the scattering amplitude with the factor $\nu_i$ as an inverse scattering length and  the threshold energy $E^{th}_i$, and the whole process represents the  transition from one amplitude $I_1= \frac{1}{\nu_1-i\sqrt{2\mu_1(E-E^{th}_1)}}$,  with the reduced mass parameter $\mu_1$ and the inverse scattering length $\nu_1$, to the channel 2 with the corresponding parameters: $I_2, E^{th}_2,\mu_2,\nu_2$, which all are calculated for the given channels. The transition probability from
the channel 1 to the channel 2 is denoted as $z(E)$, which will be the only fitting parameter of the ERM model in our analysis taking in the range $0.2-0.45$ GeV$^2$ (we suppose to compute it explicitly in the future). As a result the whole series of the transitions from 1 to 2 and back is summed up to the amplitude $f_{12}$,

\be
 f_{12}(E)= \frac{1}{1- z I_1 I_2}, I_i= \frac{1}{\nu_i- i\sqrt{2\mu_i(E-E^{th}_i)}}.
\label{6}
\ee
The values of $\mu_i$, the reduced masses of hadrons in the channel $i$,  and $\nu_i$ are found from the wave functions of
the four-quark states as it was shown in \cite{7}. The resulting values of $\nu_i$ for different channels $DD^*,D\bar D^*,...$ are given in the Appendix 1.
The form (\ref{6}) is written for the energies $E > E_1,E_2$, while for $E < E_1,E_2$ (below thresholds) the amplitude $f_1 =\frac{1}{\nu + \sqrt{2\mu_1(|E-E_1|)}}$. In the ERM it is important  that the process proceeds with the zero relative angular momentum between two mesons ($L=0$), otherwise the transition probability $z_{12}$ is much smaller and a resonance may not appear.

At this point one must specify again the recoupling mechanisms which can be formulated in two different forms:
first one is the instantaneous recoupling, where the transition from one set of the mesons to another proceeds instantaneously while the quarks keep their positions, and the transition $V(12)$ does not factorize into $V(1)V(2)$
as it was assumed in the original Recoupling Model \cite{7*}. This approach gives its definite predictions for the values of $\nu_i$, which however contradict in some cases
the properties of the resonance $T_{cc}$, and therefore we have suggested above  a more general form of the recoupling - the ERM, which is investigated below (see Appendix for the details).
In ERM we assume that the recoupling proceeds in two stages:
At first two hadrons $h_1,h_2$ collapse into one ``compound bag",  where four quarks (two quarks plus two antiquarks) are kept together by the confining interaction between all $q\bar q$ pairs. This quark compound bag has its own wave function $\Phi_i(q_1,q_2,q_3,q_4)$ and the probability amplitude of the $h_1,h_2|\Phi$ transition defines the factor $V_1(\vep_1)$, introduced above. In a similar way
the transition from the Bag state to the final hadrons $h_3,h_4$ defines the factor $V_2(\vep_2)$ and we obtain the relation:
$v_1(\vep_i)= \int d^3q_1...d^3q_4 \psi_{h_1}\psi_{h_2} \Phi_i(q_1,..q_4)$ and the same for $v_2(\vep_2)$ with the replace of $h_1,h_2$
by $h_3,h_4$. From $v_i(\vep_i)$ one defines the parameters $\nu_i$ and finally obtains the equation (\ref{6}). In \cite{7}  this procedure is described for the resonances, discussed in this paper, and in the Appendix we quote the resulting values of $\nu_i$ for different hadron systems.

Below it will be seen that presented above simple situation can become more complicated, if additional channels, beyond 1 and 2, are included and can produce additional (independent) series
of the transitions with multiple resonances (see discussion in the concluding section).

We consider now the {\bf Set A} of resonances which are presumably four-quark charmonium-like states and we give the experimental values and corresponding the ERM parameters, referring to considered resonances.\\

{\bf{Set A. {The parameters of the assumed four-quark resonances}}}

\begin{description}

\item{{\bf 1)}}~ $T^+_{cc}(3875), J^P=1^+, \Gamma= 410 ~{\rm keV}~ (T^+_{cc}) \cite{6}, [D^{*+} D^0 \to D^{*0}D^+], E_1= 3.874, E_2= 3.876 , \nu_1=\nu_2= 0.46, \mu_1= \mu_2= 0.967$ (all in GeV).
\item{\bf 2)}~ $Z_c(3900), J^P=1^+, \Gamma= 44~{\rm MeV}~\cite{2,5}, [J/\psi \rho(\omega) \to \bar D D^*]
 E_1= 3875.14, E_2= 3879 , \nu_1= 0.21, \nu_2= 0.46, \mu_1= 0.967, \mu_2= 0.623$ (all in GeV).
\item{\bf 3)}~ $ X(3930), J^P= 0^+, \Gamma= 18.8~{\rm  MeV}, \cite{5}, [J/\psi \phi \to D_s \bar D_s],
E_1= 4.12, E_2= 3.936, \mu_1= 0.767, \mu_2= 0.984, \nu_1= 0.265, \nu_2= 0.424$ (all in GeV).

\item{\bf 4)}~ $ Z_{cs}(3985), J^P= 1^+, \Gamma= 12~{\rm MeV}, \cite{3}, [D_s^- D^{*0} \to J/\psi K^*],
E_1= 3.978, E_2= 3.976, \mu_1= 0.994, \mu_2= 0.99, \nu_1= 0.41, \nu_2= 0.23$ (all in GeV).

\item{\bf 5)}~ $ X(4014), J^P=    , \Gamma= 4 \pm 11 \pm 6, \cite{5}, [D^{*+}D^{*-} \to D^{*0} \bar D^{*0}],
E_1= 4.020, E_2= 4.014, \mu_1= 1.005, \mu_2= 1.0035, \nu_1 = \nu_2= 0.44$ (all in GeV).

\item{\bf 6)}~ $ X(4140), J^P= 1^+, \Gamma= 162(40 ) ~{\rm MeV}, \cite{1,4}, [J/\psi \phi \to D_s^{*} \bar D_s^{*}],
E_1= 4.12, E_2= 4.224, \mu_1= 0.767, \mu_2= 1.056, \nu_1=0.265, \nu_2= 0.41$ (all in GeV).

\end{description}

The {\bf Set B} refers to the resonances, which include  $X(4274),X(4500),X(4700)$ and can be the candidates of high excited $c\bar c$ states, see section 4.

\section{Resonances of the Set A in the ERM formalism}

We consider two groups of the resonances: the {\bf Set A}, presumably consisting of the ground $c\bar c, q_1\bar q_2$ or
$c, c, \bar q_1, \bar q_2$ states, where $q_1, q_2$ can be $u,d,s$, and the {\bf Set B}, presumably
consisting of high $c\bar c$ excitations. Below in this section  we give the experimental values and corresponding the ERM parameters, referring to considered resonances for the {\bf Set A}.\\

\begin{description}
\item{\bf 1)}~~$ X(3875)$, or $T_{cc}^+$ \\

We start with the case 1) -- the narrow peak $X(3875)$,  discovered by the LHCb in the $D^{*}D$ system \cite{6}.
To define the structure of the cross sections we first take the recoupling parameter $z=0.2$ GeV$^2$ and then, using the parameters from the item {\bf 1},  obtain the distribution $|f_{12}(E)|^2$, which has the maximum at $E= 3876$ MeV, and the values of $|f_{12}(E)|^2$ shown below in the Table  \ref{tab.01}.
In the amplitude $f_{12}(E)$ the resulting singularity can be found in the form of  (\ref{6}) and for equal threshold masses it produces a pole nearby thresholds. Note that below in the Table~
\ref{tab.01} we increase the distance between the thresholds up to $2$ MeV to make the cross section structure more visible. It is worth to  underline that the actual singularity structure is more complicated and consists of four poles and two square-root branch points, which could contribute to the width $\sim 0.4$ MeV of the resonance, as seen from experiment \cite{6}.

Firstly, taking the recoupling parameter $z=0.2$~GeV$^2$ and putting other parameters in (\ref{6}),  we obtain the distribution $|f_{12}(E)|^2$  with the maximum at $E= 3876$ MeV and the values of $|f_{12}(E)|^2$, given below in the Table  \ref{tab.01}.

\begin{table}[h!]

\caption{The values of the $|f_{12}(E)|^2$ near the channel thresholds for the transition {\bf 1)} -- $D^{*+}D^0\rightarrow D^{*0}D^+$}

\begin{center}

\label{tab.01} \begin{tabular} {|c|c|c|c|c|c|c|} \hline

$E$~(GeV)& 3.862& 3.872& 3.874& 3.876& 3.878&3.884\\

$|f_{12}(E)|^2$& 4.39&11.9&35.6& 48.9& 10& 3.67\\

\hline

\end{tabular}

\end{center}

\end{table}

As one can see in the Table~\ref{tab.01},  the position of the resulting resonance is just at $E= 3.876$ GeV and its  width $\Gamma\approx 2$ MeV is close to the assumed distance between thresholds.
Notice that the width can be derived smaller by adjusting the value of $z$. Indeed, for $z= 0.25$ GeV$^2$ one obtains  much larger value of
$|f(E)|^2$ and the resonance  exactly at the lower threshold, $E= E_1= 3.874$ GeV, with the width around 1 MeV. Thus the varying of the parameter $z$ does not change much  resulting position
of the maximum in  $|f(E)|^2$ but can increase the resulting width of the resonance. This example shows that in the ERM a good agreement with experimental data is reached.

\item{\bf 2}~~The resonance $Z_c(3900)$ in the $\bar D^*D$ system \cite{3,4}\\

This resonance, named $Z_c(3900)$, was observed  by the BESIII in the reaction $e^+e^- \to \pi D\bar D^*$ \cite{3} and
$e^+e^- \to \pi^+\pi^- J/\psi$ \cite{4}.

Here we assume that in this system the resonance appears  due to infinite set of the recoupling transitions: $D^+\bar D^{-*} \to J/\psi \omega(\rho)$ and  consider  the string recoupling  in the framework of the ERM. The thresholds for the transition $J/\psi \rho^+ \to \bar D^0 D^{*+}$ are $E_1= 3875.14, E_2= 3879$~MeV and for the neutral channel: $\bar D^-D^{*+}$ both thresholds in channels 1 and 2 coincide at $E= 3879$~MeV. Below we consider less singular situation of noncoinciding thresholds. Here parameters $\mu_1= 0.967$~GeV, $\mu_2= 0.623$~GeV. Choosing the parameter $z= 0.1$~GeV$^2$ in (\ref{6}), one obtains the  values of the factor $|f_{12}(E)|^2$,  given  in the Table  ~\ref{tab.02}.

\begin{table}[h!]

\caption{The values of the transition probability as a function of energy in the transition {\bf 2)} }

\begin{center}

\label{tab.02} \begin{tabular} {|c|c|c|c|c|c|c|} \hline

$E$(GeV)& 3.872& 3.875& 3.877& 3.879& 3.882& 3.885\\

$|f_{12}|^2)$& 6.64& 19.76& 21.88& 27.6& 3.96& 2.34\\

\hline

\end{tabular}

\end{center}

\end{table}

The numbers from the Table 2 show the sharp peak at $E= 3879$ MeV and the width  $\sim 10$~ MeV. This result approximately
agrees with the experimental position of the resonance and its experimental width,  $\Gamma(Z_c(3900)) \approx 28$~MeV \cite{3,4}.
If one slightly varies the fitting parameter $z$, then the position of the resonance remains near the thresholds while the width can strongly increase for large values $|z-\nu^2|$.

\item{\bf 3)}~~$ X(3930)$ \cite{5} \\

This resonance was found experimentally by the Belle group in the system $\gamma + \psi(2S) \to D_s \bar D_s$ with the width $\Gamma= 22 \pm 17 \pm 4$ MeV. We consider it as a recoupling process $J/\psi \phi \to D_s \bar D_s$ with the parameters, given in the preceding section in \item{\bf 3)}. Choosing the parameter $z= 0.3$ GeV$^2$,  one can find the cross section factor $|f_{12}(E)|^2$, defined in  (\ref{1}), which is given below in Table ~ \ref{tab.03}.

\begin{table}[h!]
\caption{The values of the $|f_{12}(E)|^2$ near  the $D_s\bar D_s$ threshold}
\begin{center}
\label{tab.03}
\begin{tabular}{|c|c|c|c|c|c|c|} \hline
$E$(GeV)& 3.8& 3.85& 3.936&4.0& 4.12&4.2\\
$|f_{12}(E)|^2$&2.23& 2.75& 91.6& 2.5& 0.75&0.433\\\hline
\end{tabular}
\end{center}
\end{table}
In Table \ref{tab.03} one can see that the narrow peak is located near the lower threshold,  $E_1= 3.936$ GeV, and has the width  $\sim 30$~ MeV, which agrees with the experimental number
$\Gamma=22(17)$~MeV from \cite{5}.

\item{\bf 4)}~ $Z_{cs}(3985)$ \cite{3}\\

This resonance, observed  by BESIII in the reaction $e^+e^- \to K^+(D_s^-D^{*0})$
\cite{3}, is treated here using (\ref{6}) with parameters $\nu_1,\nu_2$, calculated as in \cite{7} and given in the Appendix and in the previous section. The results of the calculations with $z= 0.09$ GeV$^2$ are given in the Table ~\ref{tab.04}.

\begin{table}[h!]
\caption{The values of the transition probability as a function of energy in the transition {\bf 4)} }
\begin{center}
\label{tab.04} \begin{tabular} {|c|c|c|c|c|c|c|} \hline
$E$(GeV)& 3.96& 3.975& 3.98& 3.985& 3.992& 4.0\\
$|f_{12}|^2(z= 0.09)$& 2.48& 9.62& 3.95&1.87& 1.22& 1.03\\\hline
\end{tabular}
\end{center}
\end{table}
The numbers from the Table ~\ref{tab.04} show the  peak at $E= 3975$ MeV and the width  $\sim 10$~ MeV. This result
agrees with the experimental position of the resonance and its experimental width,  $\Gamma(X(3985))=(13\pm 5)$~MeV.

\item{\bf 5)} ~$ X(4014)$ \\

Here we assume that this resonance is created due to an infinite chain of the transitions $D^{*+}D^{*-} \to D^{*0}\bar D^{*0}$.
Within the ERM mechanism \cite{7} we take  the channel coupling parameter $z=0.25$, the values of $\nu_1,\nu_2$ from the Table in the Appendix and compute the corresponding values of $|f_{12}(E)|^2$ (see the Table~\ref{tab.05}).

\begin{table}[h!]
\caption{The values of the $|f_{12}(E)|^2$ near the channel thresholds for the transition {\bf 5)} }
\begin{center}
\label{tab.05} \begin{tabular} {|c|c|c|c|c|c|c|}
\hline
$E$(GeV)& 4.005& 4.010& 4.014& 4.017& 4.020&4.025\\
$|f_{12}(E)|^2$& 12.2& 29& 915& 25.3& 7.1& 6.96\\\hline
\end{tabular}
\end{center}
\end{table}
In Table~\ref{tab.05} one can see the sharp peak at $E_0= 4.014$~GeV  with the width
of  $\sim 8$ MeV, which agrees well with the experimental $\Gamma= 4 \pm 11$ MeV.

\item{\bf 6)}~~$ X(4140)$ \\

This resonance was observed by the LHCb group  in the reaction $J/\psi + \phi \to D_s^{*} + \bar D_s^{*}$  \cite{1,4}. The resulting
values of $|f_{12}(E)|^2$ for $z= 0.35$ GeV$^2$ are given in the Table~\ref{tab.06}.

\begin{table}[h!]
\caption{The values of the $|f_{12}(E)|^2$ near the channel thresholds for the transition {\bf 6)} }
\begin{center}
\label{tab.06} \begin{tabular} {|c|c|c|c|c|c|} \hline
$E$(GeV)& 4.0& 4.07& 4.12& 4.17& 4.224\\
$|f_{12}(E)|^2$& 3.4&8.67& 3.86&1.27&0.45
 \\\hline
\end{tabular}
\end{center}
\end{table}
In the Table~\ref{tab.06} one can see that the asymmetric resonance appears below the lower threshold with  $E_1(th.)= 4120$~MeV and  this effect is typical for the resonances,
produced due to the ERM mechanism. It mass $E_0\cong 4100$ is in reasonable agreement with experiment,  while the width  $\Gamma\cong 100$~MeV, is smaller than the latest experimental
number,  $\Gamma(\exp.)=162(20)$~MeV \cite{2,4}. Notice that much smaller width  $\Gamma\sim (10-15)$~ MeV was found in the earlier experiments  \cite{2,4}, which corresponds to lower values of $z$.

\end{description}

Below we summarize our results for the resonances from the Set A in comparison with data in the Table~\ref{tab.07}.

\begin{table}[h!]
\caption{The ERM results for the resonances of the Set A in comparison with experiment}
\begin{center}
\label{tab.07} \begin{tabular}{|c|c|c|c|c|c|}
\hline
$N$&Resonance&$J^P$&$\Gamma_{\exp}$MeV&$M({\rm theor})$MeV&$\Gamma({\rm theor})$MeV\\\hline
1&$X(3875)$&$1^+$&0.41 \cite{6}&3876&$<2$\\
2&$Z_c(3900)$&$1^+$&28.3 \cite{25}&3880&$10$\\
3&$X(3915)$&$0^+$&22 \cite{5}&3930&30\\
4&$Z_{cs}(3985)$&$1^+$&13 \cite{3}&3975&10\\
5&$X(4014)$&$0^+$&4-11 \cite{5}&4014&8\\
6&$X(4140$)&$1^+$&162 \cite{1,4}&4120&100\\\hline
\end{tabular}
\end{center}
\end{table}
One can see that the predictions of simple two-channel ERM, exploited above, occurs to be in  reasonable agreement with experiment.
Comparing our results with those in literature, one can notice that our conclusion of the $4q$ structure of the $X(3915), Z_{cs}(3985), X(4140)$ resonances is in agreement with  the
analysis in the papers \cite{18,19,19*}, based on the coupled channel model of $c\bar c$ and meson-meson
systems. Notice that the general structure of the channel-coupling matrix elements in both approaches is similar, however, the explicit form of the channel wave functions is different, which nevertheless leads to similar answers.

\section{The charmonium $n\,^3P_J$ states}

In this section we consider the charmonium-like states which can be associated with high excited and possibly shifted
charmonium $c\bar c$ states which can be treated in the relativistic Hamiltonian approach.
The relativistic string Hamiltonian \cite{22,23} was widely used in light, heavy-light, and heavy mesons \cite{20,21}. It acquires very simple form in the case of heavy quarkonia, since
their masses have practically no  self-energy \cite{25} and the string corrections \cite{21,26}, important for light and heavy-light mesons. In heavy-quarkonia it reduces to the
form of the spinless Salpeter equation,
\be
\left(2\sqrt{m_c^2 + \vep^2} + V_0(r)\right) \varphi_{nl}(r)=M_{\rm cog}(nl) \varphi_{nl}(r),
\label{7}
\ee
which defines the center of the gravity (the spin-averaged mass) of the $n\,^3L_J$ multiplet and the interaction $V_0$ does not include the spin-spin, the spin-orbit, and tensor potentials,
considered as a perturbation. This equation does not contain fitting parameters, namely, $m_c$ is the $c$-quark pole mass, taken here as $m_c=1.430$~GeV, which  corresponds to the conventional current mass, $\bar{m}_c(\bar{m}_c)=1.27$~GeV \cite{5*}, and the parameters of the static potential are defined on fundamental level (see below). Together with spin-dependent corrections the meson mass
can be presented as $M(J,S)= M_{\rm cog}(nL) + \delta_{\rm fs}(S,J)$, if $L\not = 0$, and $M(S,L=0) = M_{\rm cog} + \delta_{\rm ss}(S)$ for the $S$-wave states. Here $\delta_{\rm fs}(S,J)$ and $\delta_{\rm ss}(S)$ are the fine-structure and the hyperfine correction to $M_{\rm cog}$ (their parameters are discussed in \cite{21}).

At present the static potential $V_0(r)$ is known to be the sum of the gluon-exchange (GE) $V_{\rm ge}(r)$ and the confinement  $V_{c}(r)$ potentials, which parameters are
well defined at the distances $r\leq 1.2$~fm \cite{26,27}. Namely,  in the GE potential $V_{\rm ge}(r)= - \frac{4\alpha_V(r)}{3 r}$ the strong coupling is determined  by the vector  QCD constant,  $\Lambda_V(n_f=3)$, which is expressed via the conventional $\Lambda_{\overline{MS}}(n_f=3)=325(15)$~MeV, known from the high-energy experiments  \cite{5*} and the lattice studies (see discussion in
\cite{27}). It gives $\Lambda_V(n_f=3)=480(20)$~MeV. However, pure perturbative behavior of the coupling takes place only at small distances $r < 0.20$~fm , while at larger distances one needs to
use the infrared regulator $M_B$, studied in \cite{26}, where it was shown to be expressed via the string tension and equal to $M_B=1.15(10)$~GeV with $10\%$ accuracy. Taking these  values $M_B$ and
$\Lambda_V(n_f=3)$, one obtains the asymptotic (frozen) coupling $\alpha_V({\rm asym.})=0.60(3)$.

From analysis of the Regge trajectories it is known that the confinement potential $V_c(r)=\sigma_0 r$ is linear at the distances $r\la 1.2$~fm and its string tension $\sigma_0=0.181(1)$~GeV$^2$ \cite{20,26}. However, such linear potential cannot describe the masses of high excitations of light mesons \cite{26} and charmonium \cite{20}, which appear to be by $(100-200)$~MeV larger than in experiment. To explain these  mass shifts down different approaches were suggested, e.g. introducing the  flattened $V_c(r)$ potential in \cite{26}, or the screened confinement potential \cite{28}, or taking into account the meson-meson correction due to open channels \cite{29}.
In  the flattened confinement potential the  string tension $\sigma(r)$ is decreasing at large $r$ \cite{26} and later this potential was shown
to be  universal and can describe  high excitations of heavy quarkonia and heavy-light mesons \cite{20,21}. This effect -- decreasing of $\sigma(r)$ occurs due to creation of the virtual
light $q\bar q$ loops in the Wilson loop, decreasing its modulus square; this effect is not related with the threshold phenomena. However, the flattened potential contains two
phenomenological parameters, taken the same for all mesons. Notice that in contrast to the flattened potential
the parameters of the screened potential, introduced in \cite{29*}, are not universal and vary in different systems \cite{28}).

Thus we introduce flattened $V_c(r)$, but suppose that the GE potential is not screened at large distances, although this fact is not proved yet. In $V_c(r)=\sigma(r) r$
the following $\sigma(r)$ was suggested \cite{16},

\be
\sigma(r)= \sigma_0 ( 1 - \gamma f(r)),~~ \sigma_0 = 0.182~{\rm GeV}^2,
\label{8}
\ee
where
\be
f(r)= \frac{\exp(\sqrt{\sigma_0}(r - R_0))}{B + \exp(\sqrt{\sigma_0}(r - R_0))}.
\label{9}
\ee
The best fit to describe high excitations was obtained for the values $\gamma=0.40$ and $R_0= (5.5\pm 0.5)$~GeV$^{-1}$ and with this potential the masses of the $S$-wave charmonium states
were calculated in \cite{10}, together with the analysis of the Regge trajectories in charmonium. Here in Table~\ref{tab.08} we give the masses of $\chi_{cJ}$ with $n=1-5$, taking
$R_0=5.5$~GeV$^{-1}$, $B=15$, and $\gamma=0.40$. Notice that here the spin-orbit term (its perturbative part) and the tensor splitting have no fitting parameters and are defined with the same strong coupling $\alpha_V(r)$ as in the static potential.

\begin{table}[h!]
\caption{The masses of the $n\,^3P_J$ states (in MeV) in charmonium}
\begin{center}
\label{tab.08}
\begin{tabular}{|c|c|l|}
\hline
state &  $R_0=5.5$~GeV$^{-1}$  & experiment\\\hline

$1\,^3P_0$ &    3425& 3414.8(3) \\
$1\,^3P_1$ &    3505 & 3510.7(1)  \\
$1\,^3P_2$  &   3545  & 3556.2(1) \\

$2\,^3P_0$  & 3876    &  3862$^{+26}_{32}$ \cite{5*}\\
$2\,^3P_1$   & 3936  &  3871.7(2) \cite{5*}\\
$2\,^3P_2$  &  3976 &  3922.2(1.0) \cite{5*}\\

$3\,^3P_0$    & 4225  &  abs. \\
$3\,^3P_1$    &  4274  &  $ 4274^{+8}_{-6}$ \cite{5*}\\
             &               & 4294(4) \cite{2*}\\
$3\,^3P_2$  &  4304   &  abs.\\

$4\,^3P_0$  &  4493  & 4506(25) \cite{5*}\\
           &                & 4474(6)  \cite{2*}\\
$4\,^3P_1$  &  4529  &    abs. \\
$4\,^3P_2$   &  4551  &  abs. \\

$5\,^3P_0$  &   4691   &  $ 4704^{+26}_{38}$ \cite{5*}\\
             &                 & 4694(4) \cite{2*} \\
$5\,^3P_1$   & 4706    & abs.\\
$5\,^3P_2$   &  4720   & abs.\\

$6\,^3P_0$   & 4831   & abs  \\
$6\,^3P_1$   & 4856 & abs \\
$6\,^3P_2$  & 4868  & abs \\\hline
\end{tabular}
\end{center}
\end{table}

From the Table~ \ref{tab.08} one can see a good agreement with experiment for high $n\,^3P_J~(n=4,5)$ states, while the masses of
$\chi_{c1}(2P), \chi_{c2}(2P)$ occur to be larger due to large threshold effect near $D\bar D^*$ threshold.  In particular,  $M(4\,^3P_0)=4493$~MeV and
$M(5\,^3P_0)=4706$~MeV are in very good agreement with those of $X(4500)$ and $X(4700)$ \cite{4}. Also predicted mass of $\chi_{c1}(3P)$  agrees with the
mass of $X(4273)$.  Unfortunately, at present there are no experimental data on high charmoinum excitations with $J^P=1^+,2^+$ and $n=4,5$; however, these data are very important to understand the scale of the fine-structure effects of high charmonium states. These data could also provide valuable  information about behavior of the  GE potential at large distances, in particular,
the fine-structure splitting strongly decreases  for a screened GE potential. In our approach the masses of $\chi_{c1}(4P), \chi_{c1}(5P)$ are obtained to be equal to 4529~MeV and 4706~MeV,
respectively.  In conclusion we would like to note that our interpretation and the masses of $X(4274), X(4500), X(4700)$ agrees with the results in \cite{18}, where the constituent quark model with the  channel-coupling effects is used.

\section{Conclusions and an outlook}

In charmonium-like family, besides  almost pure $c\bar c$ states, like $J/\psi$ and other states below $D\bar D$ threshold, there are different types of the resonances.
Among them  the mostly $c\bar c$ states, shifted down due to strong interaction with meson-meson channels. For example,  $X(3872)$ is considered as  $c \bar c$ state with
$ J^{PC}= 1^{++}$, shifted down to the  $D\bar D^{*}$ \cite{20}. Another-- new type of the resonances- is formed, if one pair of the mesons can transform into another pair of mesons infinitely many times, what was called the recoupling mechanism \cite{7*} and studied further within  the ERM \cite{7}. Here this type of the resonances was classified as  {\bf Set A}, while   others  as {\bf Set  B}. We have demonstrated the following features of these two sets of the resonance structures:

In the case {\bf Set A } narrow peaks are possible near adjacent thresholds, which appear due to the transitions between the four-quark  (meson-meson) structures, if the  angular momentum
between the mesons is equal zero. In the case {\bf  Set B} the resonances are shifted from the positions, as predicted in the $c\bar c$ picture with the linear confinement potential, and these mass shifts down (closer to the threshold)  can reach ($100-200$)~MeV for high excitations. With the use of the flattened confinement potential with universal parameters the masses of the
$4\,^3P_0, 5\,^3P_0$ states are obtained in very good agreement with the those of the $X(4500)$ and $X(4700)$ resonances.
In our study we have identified the resonances of the  {\bf (Set A)} with the singularities, produced by the ERM mechanism, where
one pair of mesons can transfer into another pair infinitely many times, and these singularities are located near thresholds. In the ERM it is necessary that the decay thresholds should be close by
and in considered transition there are no suppression of the amplitude due to the angular momentum, or the spin-spin coefficients.  In particular,  the ERM formalism is not expected  in the case of
the $ X(4230)$ resonance with $J^P=1^-$, observed  in the $J/\psi \phi$ system. According to our analysis  in all cases {\bf (1-6)},  taking the parameters $\nu_1,\nu_2$ from the oscillator forms of the wave functions of the participating mesons, one obtains the masses and the small widths of the resonances $X(3875),Z_c(3900), X(3915),Z_{cs}(3985)$, in good agreement with experimental data. In the case of the {\bf 6)} the larger width $\Gamma \cong 100$~MeV  also agrees with data.
Summarizing, one can see that the approaches, used in our paper, help to understand the physical nature of two distinct groups of resonances in the
region (3.9--4.7)~GeV and in particular, the nature of very narrow four quark resonances. One can stress that basing on the
detailed many-channel analysis done in \cite{18,19,19*} one should include the many-channel recoupling structure for the most resonances in this region, which is planned for the future.

The authors are grateful to N. P. Igumnova for collaboration.

\section*{Appendix. {The parameters of the recoupling amplitude $f_{12}(E)$}}

 \setcounter{equation}{0} \def\theequation{A.\arabic{equation}}

The basic element in the recoupling amplitude $I_i(E)$ has the form

\be
I_i(E)= \int{\frac{d^3 p_i}{(2\pi)^3}\frac{v_i^2(p_i)}{E'(p_i) + E"(p_i) - E}},
\label{A.1}
\ee
where $v_i(p)$ is expressed as an integral of the product of four hadron wave functions participating in the recoupling process $\phi_n(k_n)$, n= 1,2,3,4. It is convenient to exploit for these wave functions the Gaussian form, e.g. $\phi_n(p)= c_n \exp(-\frac{p^2}{4\beta_n^2})$ fitted to the numerically
computed wave functions in \cite{20,21} for 15 different mesons from $J/\psi$ to $\rho$. The accuracy of these fits is better than 15 percent in the significant region where wave function is larger than 10 percent of its maximal value-see Table II in the \cite{7*}. As the result of integration over $p_n$ it was obtained \cite{7*} that $v_i^2(p_i)$ also have the
Guassian form $v_i^2(p_i)= b_i \exp(-Y_i p_i^2)$, $b_i=$const and $Y_i= F(\beta_1,\beta_2,\beta_3,\beta_4)$ . Finally, integrating over $d^3 p_i$, one obtains the final approximate form for $I_i(E)$,
used in \cite{7} and above in the paper,
\be
I_i(E)= C_i (\nu_i -i \sqrt{2 \mu_i (E- E^{th}_i)})^{-1}.
\label{A.2}
\ee
This form has an accuracy of better than 15\% for energies between two thresholds and can be used for a qualitative
analysis of the resonances in this region. The parameters is defined as $\nu_i= \frac{1}{ \sqrt{2 Y_i}}$ and are given below
in the Table \ref{tab.Ap},

\begin{table}[h!]

\caption{The values of $\nu_i$ for different charmonium-like systems  }

\begin{center}

\label{tab.Ap} \begin{tabular} {|c|c|c|c|c|c|c|} \hline

$hh$ system&$ J/\psi \omega$& $J/\psi \phi$& $DD^*,\bar DD^*$&$ D^*D^*,D^*\bar D^*$& $D_sD_s$&$D_s^*D_s^*$\\

$\nu_i$(GeV$^2)$& 0.21& 0.265& 0.46& 0.44& 0.424& 0.41\\

\hline
\end{tabular}
\end{center}
\end{table}

\end{document}